# Experimental and numerical analysis of the chromatic dispersion dependence upon the actual profile of small core microstructured fibres

**L Labonté** [1]**, P Roy** [1]**, D Pagnoux** [1]**, F Louradour** [1]**, C Restoin** [1]**, G Mélin** [2]**, and E Burov** [2]
1  XLIM, Unité Mixte CNRS-Université de Limoges n° 6172, 123 Avenue A. Thomas, 87060 Limoges Cedex, FRANCE
2  ALCATEL Resarch and Innovation, Route de Nozay, 91460 Marcoussis, FRANCE

E-mail: laurent.labonte@xlim.fr

**Abstract** : the chromatic dispersion curve of the fundamental mode in small core microstructured fibres (SCMF) is both calculated using a Finite Element Method (FEM) and measured with a low coherence interferometric method. The great sensitivity of the chromatic dispersion to variations of the geometrical parameters of SCMFs (the pitch $\Lambda$ and the diameter d) is pointed out. An excellent agreement is obtained between the numerical and the experimental results over a half micrometer spectral bandwidth [1.1 μm-1.6 μm].

## I. Introduction

Thanks to their unusual propagation properties, the microstructured fibres that guide light by Modified Total Internal Reflection (M-TIR) [1][2] have aroused a great interest in recent years. Such fibres are made of a strand of silica glass with an array of micrometer air channels running along their length. The core is constituted by one (or several) missing hole. The geometrical parameters of the fibre are the diameter of the air holes (diameter d) and the distance between the centres of adjoining holes (pitch $\Lambda$).

These fibres have successively demonstrated a wide single-mode wavelength range [3], large mode area [4] and high birefringence [5]. The chromatic dispersion, which depends on both guide and material dispersion [6], is usually dominated by the dispersion of the bulk silica in the conventional doped silica step-index fibres. On the contrary, in M-TIR, the wide choice in the geometrical parameters and the large refractive index difference between air and silica enable much greater flexibility in the design of the dispersion of the guide. Consequently, the chromatic dispersion of M-TIR can be controlled by a suitable choice of the parameters d and $\Lambda$.

In the field of non-linear (NL) optics, M-TIR with a small core are very attractive because the guided modes can be strongly confined. This results in a shift of the zero chromatic dispersion wavelength towards short wavelengths, together with an enhancement of the power density of light in the core. Such small core microstructured fibres (SCMF) are then suitable to obtain broadband continuum [7] and soliton propagation at wavelengths shorter than 1.3 μm [8]. At the opposite dual-concentric-core

M-TIRs can exhibit a highly negative dispersion @ 1550 nm that can compensate for positive dispersion of conventional fibres [9].

In the past few years, the chromatic dispersion of SCMF has been widely studied in theoretical and experimental works. For example, calculations based on the idealized cross section have been reported [10]. In this paper the geometrical parameters that are introduced in the calculations are equal to the mean parameters deduced from those measured on the SEM image of the actual cross section. Special designs for flat chromatic dispersion were proposed [11] or were performed [12]. Many applications such as telecommunications, supercontinuum generation or parametric amplification require an accurate control of the chromatic dispersion. It is well known that chromatic dispersion can be controlled thanks to parameters d and $\Lambda$ [13].

In this paper, we propose to analyse, over a broadband, the sensitivity of the chromatic dispersion to slight imperfections in actual fibre transverse profiles. Furthermore, the impact on the chromatic dispersion of d and $\Lambda$ fluctuations occurring along the fibre length is pointed out. With this aim, we compare calculations using the actual cross sections of SCMFs with broadband experimental measurements (500 nm). We calculate the chromatic dispersion of actual SCMFs using a numerical approach that is described in section II. Then, in section III, we present the experimental method that we have used to measure the chromatic dispersion over the wide spectral band. In the last part, we compare and discuss our results.

## II. Numerical modelling method for the computation of the chromatic dispersion

The chromatic dispersion D is calculated using the equation 1:

$$D = -\frac{\lambda}{c}\frac{d^2 n_e}{d\lambda^2} \qquad \text{Equation 1}$$

c is the light velocity in vacuum.

Due to the second derivative in equation 1, the spectral evolution of the effective index $n_e$ must be computed with accuracy in order to obtain a reliable computed chromatic dispersion. Many vectorial methods have already been proposed to compute the effective index into SCMFs [14]. However, most of them do not apply for actual unsymmetrical fibres [14] [15]. Among the available numerical methods, we have chosen the FEM because of its accuracy [16], but also because its principle remains valid for actual non symmetric cross section. The actual cross section of the fibre is split in elementary subspaces making a mesh. The FEM consists in solving the Maxwell equations at each nod of the mesh, taking into account the continuity conditions of the fields at the subspace boundaries. For actual fibres, an accurate description of the fibre profile can be obtained by making the mesh from SEM (Scanning Electron Microscopy) image of the fibre, even if the fibre exhibits small geometrical defects.

An example of SEM image is shown in the figure 1a. The air holes appear in dark and the silica in pale. By means of a numerical treatment that consists in comparing the darkness of each point of the SEM image to a decision level determined in the grey-scale, the accurate determination of holes boundaries is obtained (figure 1b). The whole cross section is then meshed using triangular elementary subspaces. Typically, the dimensions of these elements are chosen to be equal to $\lambda/10$ in the regions where the electric filled is expected to be high, i.e. from the centre to the first ring of holes, and it is equal to $\lambda/7$ elsewhere [16].

**Figure 1 :** (a) SEM image of the actual cross section of fibre 1, (b) holes boundaries deduced from the SEM image

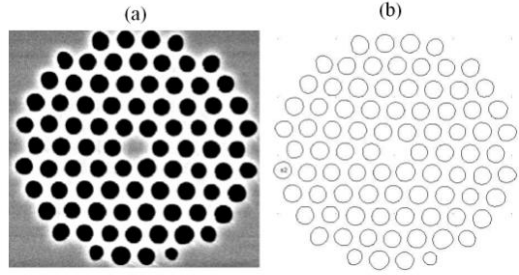

Along this process, several critical points can strongly reduce the expected accuracy on the effective index. First, the conditions to obtain the SEM image are crucial. For example, the fibre end face plane must be exactly in the SEM image plane, without residual angle. Furthermore, suitable voltage acceleration and electron beam size must be selected to avoid undesirable distortions leading to unreliable hole boundaries description on the SEM image. A second critical operation is the mesh generation. The number of elements must be compatible with the computer memory capacity while elements size must be small enough in order to accurately describe the shape of the hole of the fibre. With a suitable SEM and typical element size given before, we estimate that the uncertainty on $n_e$ value is about 0.001% leading to a chromatic dispersion uncertainty close to 5%.

**III. Experimental set-up**

To achieve the broadband chromatic dispersion measurement, the low coherence interferometry principle has been retained [17]. Our experimental set-up is based on a Michelson interferometer (figure 2). The broad band light source selected for this experiment is a supercontinuum whose spectrum is plotted in the inset of figure 2. The beam is launched into a Michelson interferometer of which one of the two arm is variable in length thanks to a moving mirror M1. A 2 cm sample of the fibre under test is positioned in the second arm. The interferometric signal is sent to a spectroscope made of a prism and a concave mirror (ensuring broad-band measurement). The channelled output spectrum is displayed on a spectral axis in the focal plane of the concave mirror. In this plane is located an InGaAs photodiode which can be translated along the spectral axis. Depending on the characteristics of the spectroscope (prism dispersion, focal length of the spectroscope mirror) and on the dimension of the sensitive area of the detector, the photodiode operates a spectral filtering into a $\delta\lambda$ spectral window around the working central wavelength $\lambda 0i$ which is a function of the detector position. On the one hand, the spectral fringes are significantly enlarged around the wavelength for which the group delays are matched in the two arms of the interferometer. On the other hand, translating the mirror M1 mainly leads to translate the spectrum along the spectral axis. Thus, for a given position of the detector (i.e. a given $\lambda 0i$ ), when the spectral fringes cross the detector area, a deep modulation of the electrical signal appears corresponding to the equality of the group delays in the two arms of the interferometer. The mirror coordinate di is measured at $\lambda 0i$ and this measurement is repeated for many $\lambda 0i$ located in the spectral band analysis. Finally the chromatic dispersion of the fibre is deduced from the formula:

$$D(\lambda_{0i}) = -\frac{1}{Lc}\frac{dd_i}{d\lambda}$$

L is the length of the sample.

Taking into account the uncertainties due to the measurements of di and to the derivative calculation, the uncertainty on the dispersion value is evaluated to be equal to ± 3 ps/(km.nm). The uncertainty on the wavelength is smaller than ± 2 nm.

**Figure 2 :** experimental setup

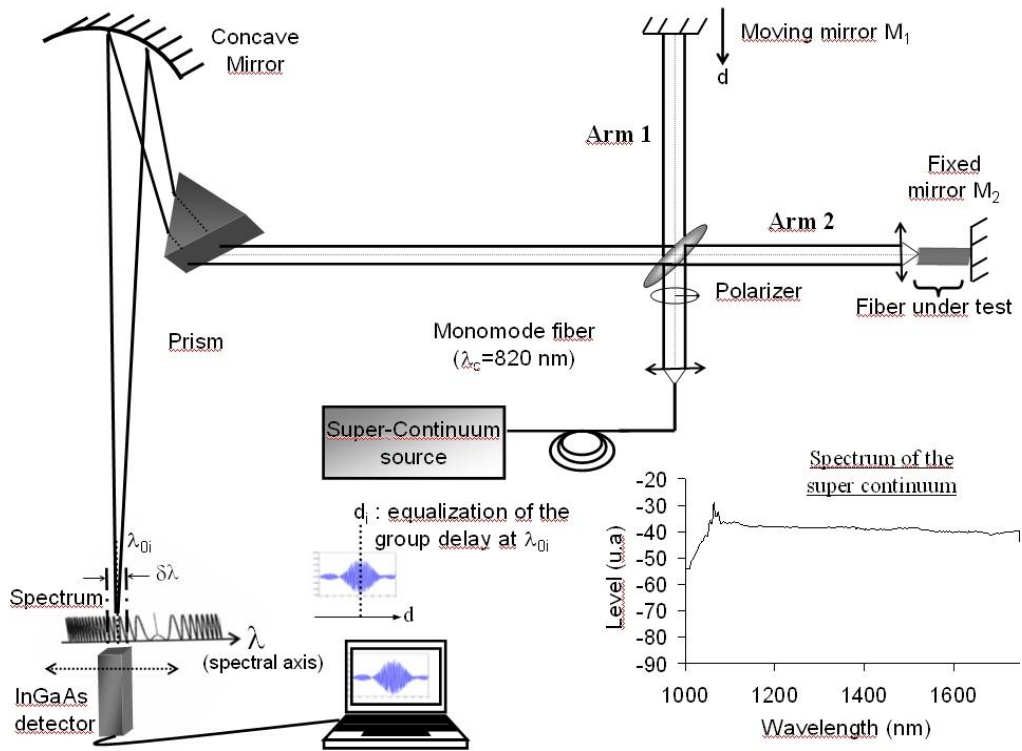

## IV. Experimental results and discussion

Three similar SCMFs with a small core and a large air-fraction cladding have been manufactured using the stack and draw technique (fibre 2 and 3 are from XLIM and fibre 1 is from Alcatel Research and Innovation). Their mean geometrical parameters d and Λ are reported in table I, together with the SEM image of their cross section. The estimated effective area of each fibre is also reported in table 1. The mean value of the effective area ($A_{eff}$) for the three fibres is close to 4 µm².

**Table 1 :** Geometrical description of the fibres

|  | Fibre 1 | Fibre 2 | Fibre 3 |
|---|---|---|---|
| Cross section (SEM Image) |  |  |  |
| d (µm) | 1.9 | 1.4 | 1.8 |
| Λ (µm) | 2.3 | 2 | 2.26 |
| d/Λ | 0.83 | 0.7 | 0.8 |
| $A_{eff}$ (µm²) @1550 nm | 3.3 | 4 | 3.9 |

We first consider the idealized cross section of fibre 1 with a perfect triangular lattice of cylindrical holes. We consider three pairs of parameters (d and Λ) close to the mean parameters of the actual fibre. The chromatic dispersion of the corresponding idealized fibres has been computed and the numerical results are shown in figure 3.

The chromatic dispersion of the fibre 1 has been measured over 500 nm and is superimposed on the same figure.

**Figure 3 :** measured and computed chromatic dispersion of the fundamental mode of the fibre

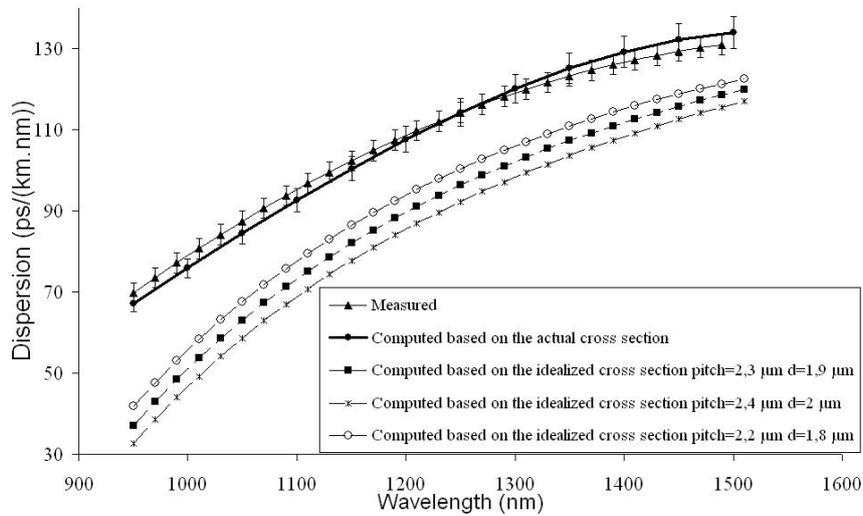

A significant discrepancy is observed between experimental result and numerical calculations obtained in the case of idealized cross sections. We can note a discrepancy up to 40 ps/(km.nm) at short wavelengths. This difference decreases as the wavelength increases but it remains larger than 10 ps/(km.nm) even at long wavelengths. Finally, we have calculated the chromatic dispersion using the actual cross section of fibre 1 deduced from its SEM image. This curve is also reported on the figure 3. In this case a very good agreement with experimental results is obtained.

The mismatch between the chromatic dispersion of actual and idealized structure is due to the fact that the considered fibre exhibits small core and a large air fraction in the cladding ($d/\Lambda > 0.6$). Indeed, as the guided mode is strongly confined into the core, a slight defect into the fibre structure (an elongated shape of the holes located around the core in the case of fibre 1) can be responsible for a large modification of the chromatic dispersion. On figure 4.a we superimpose the computed structure of fibre 1 (in grey) deduced from its SEM image and the idealized cross section (represented by black circles) which is deduced from the mean geometrical parameter of this fibre.

**Figure 4 :** a) comparison of the first ring of holes between the actual cross section of the fibre 1 and the corresponding idealized cross section b) and c) electric field orientations for the two polarisations of the fundamental mode

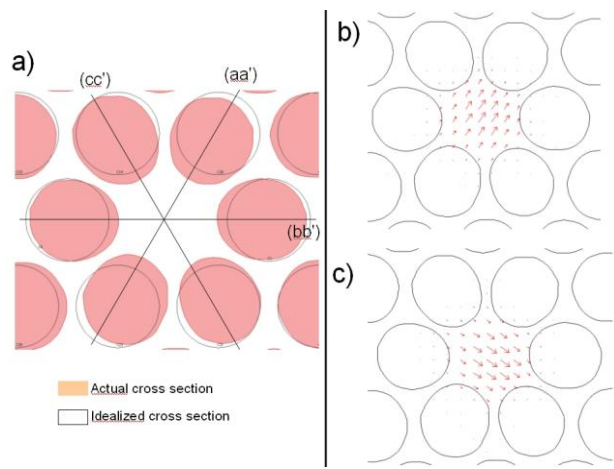

Several defects among which the shape modification and the translation toward centre are observed. On the figure 4.a, a maximum difference of around 300 nm is observed between the hole boundaries of the two structures. This clearly shows that a strong difference exists between the two structures. In particular, the translation of holes of the first ring toward the centre of the fibre decreases the area of the core. It appears that this has the same influence as a decrease of the pitch of an idealized cross section on the chromatic dispersion. It has been demonstrated that a decrease of the pitch induces an increase of the chromatic dispersion [18], which explain the mismatch observed between idealized and actual cross section on the figure 3.

Furthermore, one can notice the maximum difference of around 300 nm is observed on the (aa') axis (figure 4.a). On (bb') and (cc') axis, the mean differences are respectively evaluated to 160 nm and 170 nm. Figure 4.b and 4.c, illustrating the electric field orientations of the two polarization modes, show that the major defect, which is observed on (aa') axis, determines one neutral axis of the fibre. It is well known that the effective index difference of the two polarization modes (i.e. phase birefringence) is strongly dependent on cross section geometry [19]. Consequently, this paper now investigates the chromatic dispersion of the two polarization modes.

The computed chromatic dispersion curves of the two polarization modes of fibre 2, which exhibits a high phase birefringence equal to 1.24e-4 @ 1550 nm [19], are plotted in figure 5. To compare these numerical results with experimental ones, the measurement of the chromatic dispersion of the two polarizations has been performed. The broadband radiation that enters in the interferometer is linearly polarized thanks to a broadband polarizer located at the input of the experimental setup (figure 1). The neutral axes of the tested sample are aligned with the polarizer axes. Taking into account the experimental uncertainties, we cannot distinguish the two experimental curves referring to the polarization modes. Consequently, only one measured chromatic dispersion curve of fibre 2 is reported on the figure 5. The maximum value is around 1350 nm and is equal to 115 ps/(km.nm). The two numerical curves and the experimental one are in good agreement. In the case of this fibre, the geometrical defects that are at the origin of the birefringence do not introduce a measurable differential chromatic dispersion between the two polarization modes.

**Figure 5 :** measured and computed chromatic dispersion of the polarization modes of the fibre 2

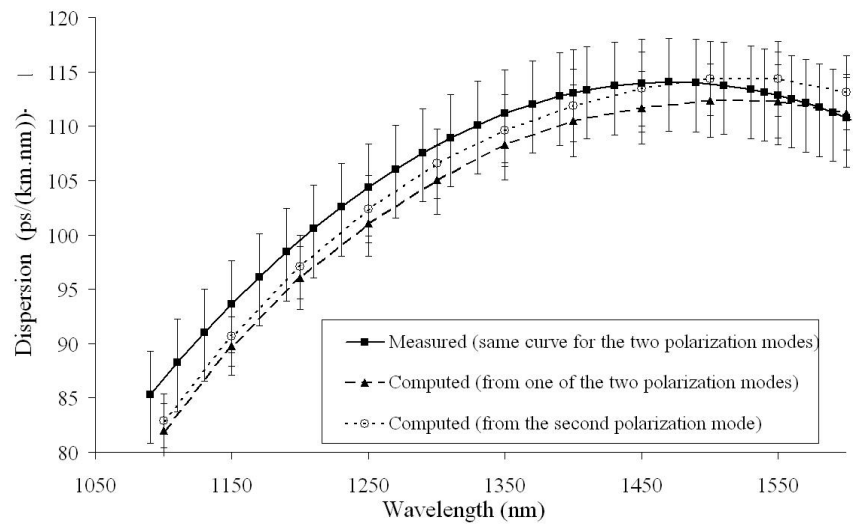

Finally the chromatic dispersion of different short length samples of the same fibre have been measured and compared. Figure 6 shows both experimental and theoretical results obtained with two 10 m away samples of fibre 3. The length of these samples was equal to 2 cm. For each sample, a difference lower than 5 ps/km.nm is obtained between numerical and experimental curves. However, the chromatic dispersion curves exhibit large discrepancies between the two samples. The first one increases from 1150 nm to 1600 nm whereas the second one has a maximum at 1370 nm. More than

30 ps/(km.nm) difference can be measured at 1150 and 1550 nm. A careful comparison between the geometry of the cross section of the two samples is achieved thanks the superimposition of the two structures deduced from the two SEM images of the samples (figure 7).

**Figure 6 :** measured and computed chromatic dispersion of the polarization modes of the fibre 3

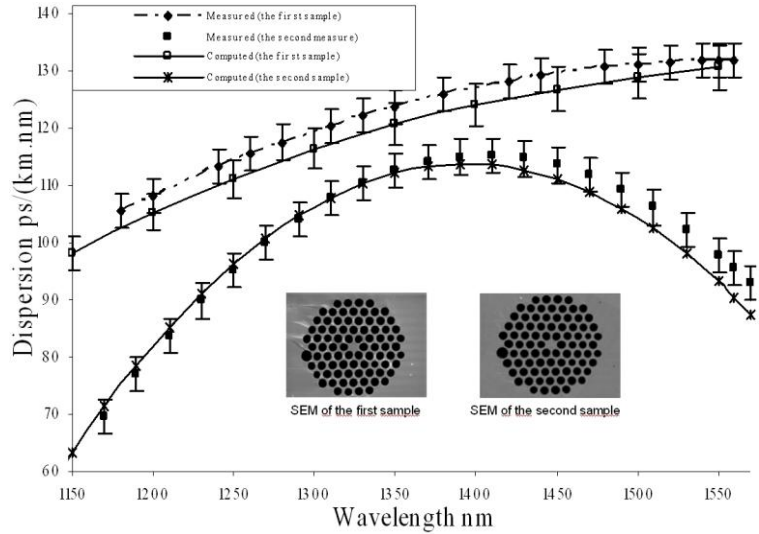

The maximum hole boundary discrepancy is smaller than 100 nm (figure 7). As this discrepancy is smaller than 3% of core diameter, this result could appear unexpected without taking into account the high sensitivity of the chromatic dispersion to the d and $\Lambda$ parameters. This experiment clearly demonstrates that slight fluctuations of the parameters along the fibre under test can have a great influence of the chromatic dispersion on the fibre. Such a high sensitivity is mainly due to the strong interaction between the guided field and the first ring of holes which diameters or position can change along the fibre. Previous numerical studies show that this sensitivity decreases as the size of the core increases [19]. Up to now, from a technological point of view, bringing under control d and $\Lambda$ over long length of fibre with accuracy better than few 10 nm seems to be very challenging. Consequently, the chromatic dispersion cannot be considered as constant along the fibre.

**Figure 7 :** comparison of the geometry of the two samples

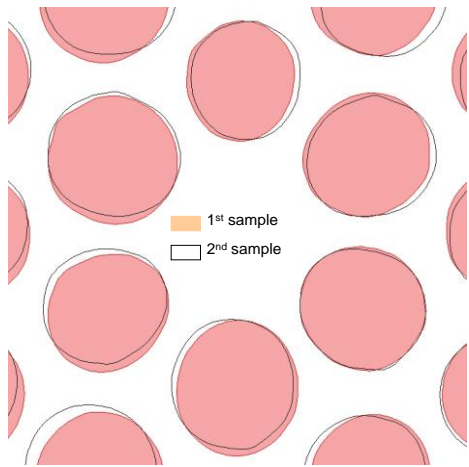

## V. Conclusion
In this paper, we report both numerical and experimental analysis of the chromatic dispersion of three actual SCMFs, with d and $\Lambda$ respectively smaller than 1.9 µm and 2.3 µm. On the one hand, the chromatic dispersion of 2 cm long samples of fibre was measured over a 500 nm spectral bandwidth thanks to a Michelson interferometer. On the other hand, the chromatic dispersion of these samples was calculated starting from the spectral evolution of the effective index of the fundamental mode computed by means of a FEM method. We have shown that, in order to obtain a reliable evaluation of the chromatic dispersion, the actual cross section must absolutely be considered instead of the idealized cross section. Furthermore, for the fibre 2 (d=1.4 µm and Λ=2 µm), we have pointed out that the chromatic dispersion of the two polarization modes are very close whereas fibres exhibit a strong birefringence due to slight geometrical defects.

Finally, we have shown the strong sensitivity of the chromatic dispersion on the parameters d and Λ. The fabrication of SCMFs requires a crucial control of the size of the holes of the first ring. For example, for the fibre 3 which parameters are equal to d=1.8 µm and Λ=2.26 µm, we have shown that a hole size variation lower than 10 nm induces a significant modification of the chromatic dispersion curve. Finally, we have considered two 2 cm long samples 10 m spaced. Both measured and calculated chromatic dispersion have led to demonstrate that the chromatic dispersion can widely vary along the fibre due to slight hole diameter or position fluctuations occurring during the manufacturing process.